\documentclass[review,doublespacing]{elsart}
\usepackage{graphicx,natbib,amssymb}

\journal{New Astronomy}
\begin{document}
\begin{frontmatter}

\title{Bimodal Distribution of Area-Weighted Latitude of Sunspots And Solar North-South Asymmetry}
\author[Chang]{Heon-Young Chang\corauthref{cor}},
\corauth[cor]{Corresponding author.} \ead{hyc@knu.ac.kr}

\address[Chang]{Department of Astronomy and Atmospheric Sciences,\\
Kyungpook National University,\\ 1370 Sankyuk-dong, Buk-gu, Daegu
702-701, Korea}

\begin{abstract}
We study the latitudinal distribution of sunspots observed from 1874 to 2009 using the center-of-latitude (COL). We calculate COL by taking the area-weighted mean latitude of sunspots for each calendar month. We then form the latitudinal distribution of COL for the sunspots appearing in the northern and southern hemispheres separately, and in both hemispheres with unsigned and signed latitudes, respectively. We repeat the analysis with subsets which are divided based on the criterion of  which hemisphere is dominant for a given solar cycle. Our primary findings are as follows: (1) COL is not monotonically decreasing with time in each cycle. Small humps can be seen (or short plateaus) around every solar maxima. (2) The distribution of COL resulting from each hemisphere is bimodal, which can well be represented by the double Gaussian function. (3) As far as the primary component of the double Gaussian function is concerned, for a given data subset, the distributions due to the sunspots appearing in two different hemispheres are alike. Regardless of which hemisphere is magnetically dominant, the primary component of the double Gaussian function seems relatively unchanged. (4) When the northern (southern) hemisphere is dominant the width of the secondary component of the double Gaussian function in the northern (southern) hemisphere case is about twice as wide as that in the southern (northern) hemisphere. (5) For the distribution of the COL averaged with signed latitude, whose distribution is basically described by a single Gaussian function, it is shifted to the positive (negative) side when the northern (southern) hemisphere is dominant. Finally, we conclude by briefly discussing the implications of these findings on the variations in the solar activity.
\end{abstract}
\begin{keyword}
Sun: sunspots, Sun: activity, Methods: data analysis
\end{keyword}
\end{frontmatter}

\section{Introduction}

Since sunspots can be considered as tracers of freshly emerged magnetic flux, any models attempting to predict the magnetic flux large-scale distribution should be able to explain  the temporal and spatial distributions of the sunspots (e.g., Parker 1955; Babcock 1961; Leighton 1969). The cyclic behavior of sunspot number (area) in time has been well-known for a long time. The variation of the sunspot number (area) with time, or  solar cycle, could be studied on the basis of the variation of the latitudinal position of the sunspots with time frequently described by the 'butterfly diagram'. For, sunspot latitude is an indication of the phase of the sunspot cycle: within a cycle, the higher the latitude, the earlier the phase in the cycle. The first version of such a diagram displaying the sunspot group distribution in latitude for each synodic rotation of the Sun was presented by Maunder (1904). One aspect which complicates the comparison of a theoretical model with the observed butterfly diagram is  that the diagram takes no account of the sunspot lifetime, or the spatial size. Since all sunspot groups are given equal weight, regardless of their temporal and spatial extention, the diagram is dominated by small sunspots which scatter over wider ranges than the larger ones. For instance, the smallest 65\% of the sunspots contribute only up to $\approx 10 \%$ of the
total spotted area (Ternullo 2007b). In addition to this, another difficulty is that the mean latitude of sunspot groups, provided that each group is assigned a weight equal to its area, does not steadily drift toward the equator (Ternullo 1997, 2001, 2007a, 2010; Norton \& Gilman 2004), at variance with the picture commonly assumed in theoretical calculations.

There are some efforts on finding the sunspot formation latitude zone to make the butterfly diagram more sensible. For instance, it has been reported that spots are not uniformly distributed in the spot zone, but repeatedly appear in some special, relatively small photospheric portions (active regions), tightly limited in latitude. The emergence of new active regions mimics an abrupt shift - either equatorward or poleward - of the spot occurrence site (Li et al. 2003; Solanki, Wenzler, \& Schmitt 2008; Ternullo 2010). The drift of the centroid of the sunspot area toward the equator in each hemisphere has also been examined (Li, Yun, \& Gu 2001; Li et al. 2002a; Hathaway et al. 2003). It is found that the drift rate slows as the centroid approaches the equator. A correlation between the solar-activity level and some characteristics of the latitude distribution of the sunspots has been studied and exercised to predict the solar activity level (e.g., Kane 2007; Javaraiah 2008; Miletskii \& Ivanov 2009). Furthermore, there is also some theoretical research employing such a latitudinal distribution in developing their model (e.g., Pelt et al. 2000; Solanki, Wenzler, \& Schmitt 2008; Goel \& Choudhuri 2009).

Since it is found that there is a systematic time lag or lead of a few years between the northern- and southern-hemispheric solar activity in a cycle, the North-South asymmetry of solar activity has been related to phase shifts between a pair of wings in the butterfly diagram (Waldmeier 1957, 1971; Pulkkinen et al. 1999; Li et al. 2002b; Temmer, Veronig, \& Hanslmeier 2002; Temmer et al. 2006; Zolotova \& Ponyavin 2006, 2007; Donner \& Thiel 2007; Li et al. 2008; Zolotova et al. 2009; Li, Gao, \& Zhan 2009a,b; Li, Liang, \& Feng 2010; Li et al. 2010). The solar North-South asymmetry is considered as one of the most interesting properties of solar activity. Asymmetries between the northern and southern hemispheres have been found in various solar indices (Roy 1977;  White \& Trotter 1977; Swinson, Koyama, \& Saito 1986; Vizoso \& Ballester 1990; Schlamminger 1991;  Carbonell, Oliver, \& Ballester 1993; Verma 1993; Oliver \& Ballester 1994; Krivova \& Solanki 2002;   Temmer, Veronig, \& Hanslmeier 2002; Ballester, Oliver, \& Carbonell 2005; Berdyugina \& Usoskin 2003;  Verma 1987; Ata\c{c} \& \"{O}zg\"{u}\c{c} 1996, 2001;  \"{O}zg\"{u}\c{c} \& \"{U}cer 1987; Tritakis, Petropoulos, \&  Mavromichalaki 1988; Vizoso \& Ballester 1989; Gigolashvili et al.  2005; Antonucci, Hoeksema, \& Scherrer 1990; Mouradian \& Soru-Escaut 1991; Knaack, Stenflo, \& Berdyugina 2004, 2005; Javaraiah \& Gokhale 1997;  Gigolashvili, Japaridze, \& Kukhianidze 2005; Javaraiah \& Ulrich 2006; Zaatri et al. 2006; Chang 2007, 2008, 2009).

In this paper, we investigate the distribution of the area-weighted latitude of sunspots. Earlier studies on the butterfly diagram have been carried out with daily values and mainly focused on either the time evolution in a cycle (e.g., Solanki et al. 2000) or the variations of characteristics of the latitudinal distribution over solar cycles (Solanki et al. 2000; Solanki, Wenzler, \& Schmitt 2008). Instead, we take the monthly average from the data set over a much longer period than one solar cycle to increase the statistical significance. Study for the bimodality and its statistical characteristics in the case when the monthly average has been carried out over each individual cycle has been carried out and discussed in elsewhere (e.g., Chang 2011). We employ the double Gaussian function rather than a single function to describe the distribution of the area-weighted latitudes where sunspots form (cf. Li et al. 2003). We address a question of whether two distributions resulted from sunspots in the northern and southern hemispheres are compatible. Upon answering to this question we are further led to address another interesting issue which is related to the solar North-South asymmetry. In other words, we further compare the distributions resulting from the subsets divided by the criterion of which hemisphere dominates in that particular solar cycle. That is, by using the double Gaussian function, we explore in details the characteristics of the solar North-South asymmetry in terms of the latitudinal distribution of sunspots.

In \S 2, we begin with brief descriptions of data and how the center-of-latitude is defined. In \S 3, we present results of the distribution of the center-of-latitude and discuss its implications on the North-South asymmetry of solar activity. In \S 4, we conclude with a summary and discussion of our results.

\section{Data}

We have used for the present analysis the Greenwich sunspot group data during the period from 1874 to 1976, and the sunspot group data from the Solar Optical Observing Network (SOON) of the US Air Force (USAF)/ US National Oceanic and Atmospheric Administration (NOAA) during the period from 1977 to 2009. The daily sunspot area and its latitude have been taken from the NASA website\footnote{${\rm http://solarscience.msfc.nasa.gov/greenwch.shtml}$}. These derived data include the correction factor of 1.4 for data after 1976 (e.g., Javaraiah, 2007).

\section{Center-of-Latitude of Sunspots}

In Figure 1 we show the 'center-of-latitude (COL)' as a function of time during the period from 1874 to 2009, which is defined by
\begin{equation}
{\rm COL}= \sum_i (a_i \times L_i)/\sum_i a_i,
\end{equation}
where $a_i$ is the $i$-th sunspot group area and $L_i$ is its latitude (Cho \& Chang 2011). The summation is carried out over a calendar year and a calendar month, instead of the Carrington month. Thick and thin curves represent the yearly averaged COL and the monthly averaged COL, respectively. In the upper panel, we separately calculate and plot the COL for sunspots appearing in the northern and southern hemispheres. The positive and negative signs represent the northern and southern hemisphere, respectively. Obviously, the solar cycle begins when COL is high in general and that COL essentially decreases as the solar cycle proceeds. We stress, however, that COL is not monotonically decreasing as commonly assumed, which can be clearly seen with the yearly averaged COL (cf. Ternullo 2007a). Small humps (or short plateaus) between solar minima are seen commonly at a  latitude of $\sim \pm 10^\circ$, which can be found in plots of similar works (Antalova \& Gnevushev 1983; Li et al. 2003; Ternullo 2007a). Humps appear in fact around every solar maxima over time. They may well be related to the latitudes of higher activity, or 'active latitude' (cf. Solanki, Wenzler, \& Schmitt 2008). In the middle panel, the averaged COL for sunspots appearing in both hemispheres is plotted taking into account the sign of the latitudes. If sunspots are randomly distributed along the mean latitude of the butterfly diagram, which is also assumed as symmetrical with respect to the equator, COL should be in a form of random noise. What is seen, however, is that the difference is large at around the solar minima and far from the random distribution. Consequently, the wings of the butterfly diagram are not symmetrical, or steadily drifts to the equator. In the bottom panel, we plot the averaged COL for sunspots appearing in two hemispheres, but the latitude is considered as unsigned, which is thus defined by
\begin{equation}
{\rm COL}= \sum_i(a_i \times |L_i|)/\sum_i a_i,
\end{equation}
such that the absolute value of the latitudes is used rather than the signed latitudes itself. Small humps seen in the upper panel are seen more definitely. It should be pointed out, therefore,  that COL is not described by a second order polynomial curve. It may require a higher order polynomial curve to fit the mean latitude of the sunspot activity as adopted to describe the latitude migration of the solar filaments (Li 2010). It may also be interesting to ask whether it should be regarded as the essential features of the 11-year cycle.

\section{Latitudinal Distribution of Sunspots}

In Figure 2, we show histograms of COL and their best fits using the double Gaussian function,
which is given by $A_1 \exp(-[x-x_1]^2/2\sigma_1^2)+A_2 \exp(-[x-x_2]^2/2\sigma_2^2)$, where $A_1$  and $A_2$, $x_1$  and $x_2$, $\sigma_1$  and $\sigma_2$, represent the amplitude of each component, the central value, the width, respectively. All the histograms in the following result from monthly averaged COLs. We count the number of COL into bins with a width of  $1^\circ$ for the period from 1874 to 2009, in order to be independent of a particular solar cycle and to increase the statistical significance. We then fit a double Gaussian function to the histogram by adjusting 6 parameters (two amplitudes, two central values, and  two widths) at the same time. Solid, dashed, and dotted curves represent the observed histogram, the best fit of the double Gaussian function, and the individual Gaussian function, respectively. In the upper left and right panels, results of the histograms from the sunspots in the northern and southern hemispheres are shown, respectively. In the lower left panel, histograms are obtained by sunspots in both hemispheres using unsigned latitudes of the sunspots. In the lower right panel, on the other hand, histograms are obtained by sunspots in both hemispheres using the signed latitudes of the sunspots. In other words, histograms in the bottom row from left to right are a result of the last and middle panels in Figure 1, respectively.

As shown in Figure 2, the distribution of COL as a result from each hemisphere is bimodal, which is well represented by the double Gaussian function having maxima of $\sim 11^\circ$ and $\sim 20^\circ$. Of interest is that the profiles are skewed such that the main component is closer to the equator with the secondary component towards the higher latitudes. Apparently, the width of the fitting function is somewhat wider for the northern hemisphere. We perform the Student's t-test with the assumption that two data sets have different variances (see Press, Teukolsky, \& Vetterling 2007). We have calculated the probability that the distributions of COL from the two hemispheres have significantly different means and found it insignificant. That is, the probability we obtain is 19.21 \%. Hence, one cannot accept the hypothesis that the COL distributions from the two hemispheres have different means. Not surprisingly, the unsigned average of the COL from the two hemispheres is bimodal too, whose best fit is peaked at $\sim 11^\circ$ and $\sim 20^\circ$. On the other hand, the signed average of COL from two hemispheres basically results in a single Gaussian function peaked at $\sim 0^\circ$. The obtained fitting parameters are listed in Table 1.

In Figures 3 and 4, we show histograms of COL from two subsets and their best fits using the double Gaussian function. Here we divide the whole data set into two subsets according to which hemisphere is dominant in the solar North-South asymmetry for a given solar cycle. Figures 3  and 4 are a result of the sunspots during the period that the northern and southern  hemispheres are magnetically dominant, respectively. The northern hemisphere has been dominant during the cycles from 14 to 20, and the southern hemisphere in cycles 12, 13, 21, 22, and 23 (e.g., Li et al. 2003; Chang 2008). Line types in Figures 3 and 4 are same as in Figure 2. Here one may find similar key features as seen in Figure 2, such as, bimodality. When comparing Figures 3 and 4, however, several interesting points can be noted, and they are as follows:

(1) As long as the main component of the double Gaussian function at lower latitude is concerned, regardless of which hemisphere is magnetically dominant, the main component of the double Gaussian function seems relatively unchanged in the central position and in the full-width-at-half-maximum (FWHM), except for the amplitude. The amplitude is affected by the number of solar cycles simply because the number of solar cycles when the northern hemisphere is dominant is 7, yet the number of solar cycles when the southern-hemisphere is dominant is 5. They are all centered at  $\sim 12^\circ$ with FWHM of $\sim 6^\circ$. Naturally, the distribution of COL of the sunspots appearing in both hemispheres with unsigned latitudes is also quite similar from one hemispheric dominance to another.

(2) On the other hand, the secondary component of the double Gaussian function at higher latitudes seems to have a role in the sense that even though their central position remains fixed at $\sim 21^\circ$ or  $\sim 22^\circ$, their width varies systematically. That is, when the northern (southern) hemisphere is dominant the width of the secondary component in the northern (southern) hemisphere case is about twice as wide as that in the southern (northern) hemisphere, as shown in Tables 2 and 3. It may imply that when a particular hemisphere is dominant sunspots in that hemisphere emerge at higher latitude than the other hemisphere. In addition to this, the total width of the double Gaussian function (for example, $\sigma_1+\sigma_2$) due to the northern (southern) hemisphere is a little bit wider when the northern (southern) hemisphere is dominant, even though it is less obvious than what we mentioned previously. This also may tell us that the sunspots not only emerges at higher latitudes but also ends at somewhat lower latitudes in the hemisphere than in the other one when a particular hemisphere is dominant.

(3) For the distribution of COL averaged with the signed latitudes, a little population can be found in the higher latitudes in the north (south) of the equator rather than in the south (north) of the equator when the northern (southern) hemisphere is dominant. This may reflect the fact that the secondary component in the distribution of a hemisphere is wider than that of the other hemisphere when that particular hemisphere is dominant. Nonetheless, regardless of which hemisphere is magnetically dominant, the distribution is basically described by a single Gaussian function as seen in Figure 2. Another interesting point to make here is that the distribution is shifted to the positive (negative) side when the northern (southern) hemisphere is dominant (see Tables 2 and 3).

\section{Summary And Conclusion}

The distribution of area-weighted latitude of the sunspots appearing during the period from 1874 to 2009 has been studied. We first determine the center-of-latitude (COL) by averaging the latitude with the weight function in area. Then the latitudinal distribution of COL was formed for sunspots appearing in the northern and southern hemispheres separately, and in both hemispheres with unsigned and signed latitudes, respectively. We further obtain the best fit of the observed distribution with the double Gaussian function and compare results. We repeat the same analysis with sunspots appearing either when the northern hemisphere is dominant or when the southern hemisphere is dominant. The most interesting result of our analysis concerns the north-south asymmetry of the solar activity. The solar North-South asymmetry can be traced in the distributions of COL.

Our main findings are as follows:

(1) COL is not monotonically decreasing as commonly assumed. Small humps (or short plateaus) between solar minima can commonly be seen at  latitudes of $\sim 10^\circ$. Humps appear in fact around every solar maxima. A higher order polynomial curve may be required to fit the mean latitude of the sunspot activity.

(2) The distribution of COL resulted from each hemisphere is bimodal, which is well represented by the double Gaussian function having maxima at $\sim 11^\circ$ and $\sim 20^\circ$.  One cannot rule out that the hypothesis that the COL distributions from the two hemispheres have same means. The signed average of the COL from the two hemispheres basically results in a single Gaussian function peaked at $\sim 0^\circ$.

(3) As far as the main component of the double Gaussian function is concerned, for a given data subset, the distributions due to the sunspots appearing in two different hemispheres are alike. Regardless of which hemisphere is magnetically dominant, the main component of the double Gaussian function seems relatively unchanged in the central position and in the FWHM, except for the amplitude.

(4) On the other hand, when the northern (southern) hemisphere is dominant the width of the secondary component in the case of the northern (southern) hemisphere is about twice as wide as that in the southern (northern) hemisphere. Thus, one may wish to determine which hemisphere is dominant by measuring the width of the secondary component at higher latitudes, which is indeed a result of the sunspots appearing in an earlier phase of the solar cycle.

(5) For the distribution of COL averaged with signed latitude, a little population can be found in the higher latitudes in the north (south) of the equator rather  than in the south (north) of the equator  when the northern (southern) hemisphere is dominant. Regardless of which hemisphere is magnetically dominant, the distribution is described by a single Gaussian function, which is shifted to the positive (negative) side when the northern (southern) hemisphere is dominant.

We conclude by pointing out that links between what we have found here and shorter periodic phenomena than $sim$ 11 years, e.g., those of $sim$ 2 year periods, should be worth further exploration. Examples of related issues are as follows: First, the non-monotonic behavior of the COL may be related to shorter-periodic oscillations of the sunspot cycles rather than 11 years (e.g., Ternullo 2007a). On the basis of the reasonable assumption that, the faster the tachocline rotation is, the more efficient the magnetic flux production is, it seems reasonable that the tachocline rotation-rate oscillations may affect the photospheric activity, or sunspot formation. In fact, Krivova \& Solanki (2002) attempted to demonstrate that the photospheric activity can trace the tachocline rotation-rate oscillations that helioseismic data have revealed (Howe et al. 2000). Secondly, the latitude of greatest activity is found to be associated with the boundary zone between high and low rotation rates, where the shear is at its maximum (e.g., Howard \& LaBonte 1980). This pattern, referred to as torsional oscillations, is so far known as symmetric about the equator (Howe 2009 and references therein). The asymmetric behavior of COL may be the first signature that torsional oscillations and thus deep-seated circulation patterns are not symmetrical with respect to the equator.

\section*{Acknowledgements}
The author is grateful to Sasha Kosovichev and Philip Scherrer for useful discussions and hospitality while visiting the Hansen Experimental Physics Laboratory at Stanford University where this work has been done, and sincerely thanks to anonymous reviewers for critical comments which greatly improve the original version of the manuscript. This work was supported by the National Research Foundation of Korea Grant funded by the Korean Government(NRF-2010-013-C00017).

\clearpage

\begin{table}
\begin{center}
\caption{The 6 parameters obtained from fitting the double Gaussian function to the distributions of the COL as a result from the data set of sunspots observed from 1874 to 2009. ${\rm A_k}$, $ \overline{{\rm COL}}_k$, and $\sigma_k$ are the amplitude, the central position, and FWHM of the individual Gaussian profile, respectively, and subscripts 1 and 2 stand for the primary and the secondary components of the double Gaussian function. In the first and second rows, we show fitting results from the distributions of COL of sunspots appearing in the northern and southern hemispheres, respectively. In the third and fourth rows, we show fitting results from the distributions of COL  of the sunspots appearing in both hemispheres with unsigned and signed latitudes, respectively.}
\vspace{2mm}
\begin{tabular}{lcccccc}
\hline\hline
& ${\rm A_1}$   &$ \overline{{\rm COL}}_1$ &$\sigma_1$& ${\rm A_2}$   &$ \overline{{\rm COL}}_2$ &$\sigma_2$\\ \hline
Nor. Hem.&114.8 & 11.8 & 6.4     & 32.5 & 21.0 & 7.6 \\
Sou. Hem.&105.2 &10.9  & 5.3     & 46.0 & 20.0 & 6.2\\
Unsigned &129.6 &11.2  & 4.9     & 50.9 &20.4  & 5.2\\
Signed   &87.02  &-0.1  & 10.0    & 4.6  &23.1  & 7.4\\
\hline \hline
\end{tabular}
\end{center}
\end{table}

\begin{table}
\begin{center}
\caption{Similar to Table 1, but obtained from the distributions of COL as a result from the data subset of sunspots appearing in the solar cycles when the northern hemisphere is dominant.}
\vspace{2mm}
\begin{tabular}{lcccccc}
\hline\hline
&${\rm A_1}$   &$ \overline{{\rm COL}}_1$ &$\sigma_1$& ${\rm A_2}$   &$ \overline{{\rm COL}}_2$ &$\sigma_2$\\ \hline
Nor. Hem.&62.2   &11.7  &5.8      &20.8   &21.5  &5.9 \\
Sou. Hem.&63.6   &11.8  &6.0      &28.8   &22.1  &3.0 \\
Unsigned &71.4   &11.6  &5.1      &25.5   &20.9  &5.0 \\
Signed   &49.0   &1.1   &9.6      &-0.1   &-3.4  &1.5 \\
\hline \hline
\end{tabular}
\end{center}
\end{table}

\begin{table}
\begin{center}
\caption{Similar to Table 1, but obtained from the distributions of COL as a result from the data subset of sunspots appearing in the solar cycles when the southern hemisphere is dominant.}
\vspace{2mm}
\begin{tabular}{lcccccc}
\hline\hline
&${\rm A_1}$   &$ \overline{{\rm COL}}_1$ &$\sigma_1$& ${\rm A_2}$   &$ \overline{{\rm COL}}_2$ &$\sigma_2$\\ \hline
Nor. Hem.&43.4   &12.0  &6.8      &17.5   &21.0  &2.8 \\
Sou. Hem.&42.9   &11.6  &6.0      &17.5   &21.2  &6.1 \\
Unsigned &51.6   &11.6  &5.3      &22.6   &20.8  &3.9 \\
Signed   &37.5   &-1.8  &9.1      &0.5   &21.3  &14.2 \\
\hline \hline
\end{tabular}
\end{center}
\end{table}

\clearpage

\begin{figure}
\resizebox{\hsize}{!}{\includegraphics{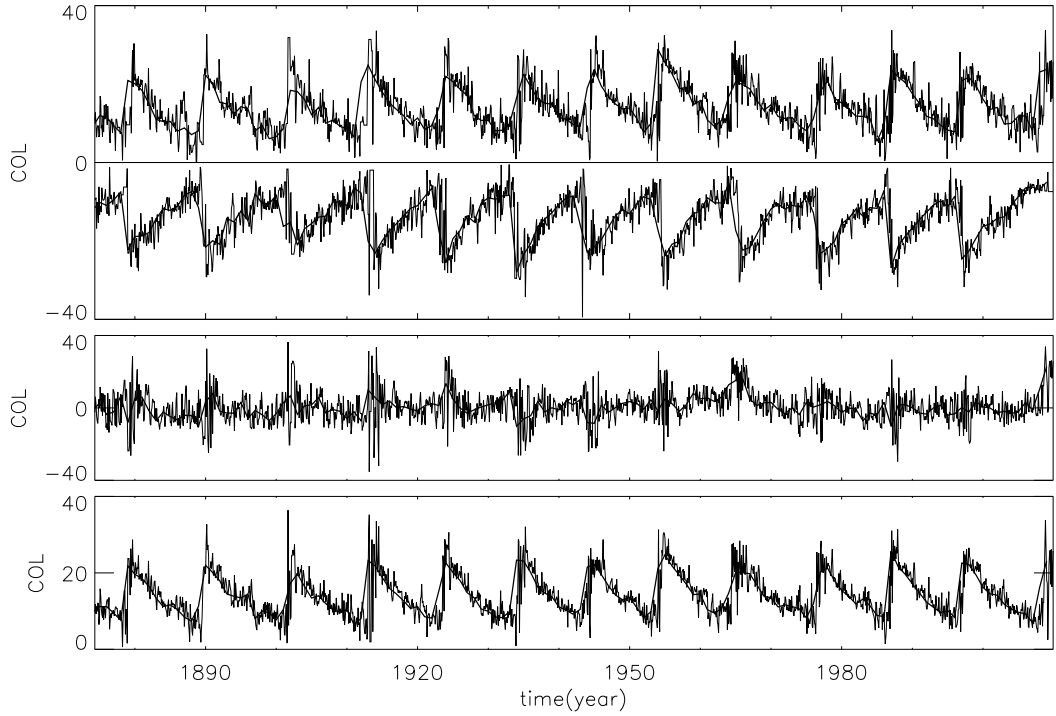}} \caption{'Center-of-latitude (COL)' as a function of time during the period from 1874 to 2009. Thick and thin curves represent the yearly averaged COL and the monthly averaged COL, respectively. In the upper panel, we separately plot COL for the sunspots appearing in the northern and southern hemispheres. In the middle panel, the averaged COL for sunspots appearing in both hemispheres is plotted taking into account its sign. In the bottom panel, we plot the averaged COL for sunspots appearing in two hemispheres but the latitude is considered as unsigned.}
\end{figure}

\begin{figure}
\resizebox{\hsize}{!}{\includegraphics{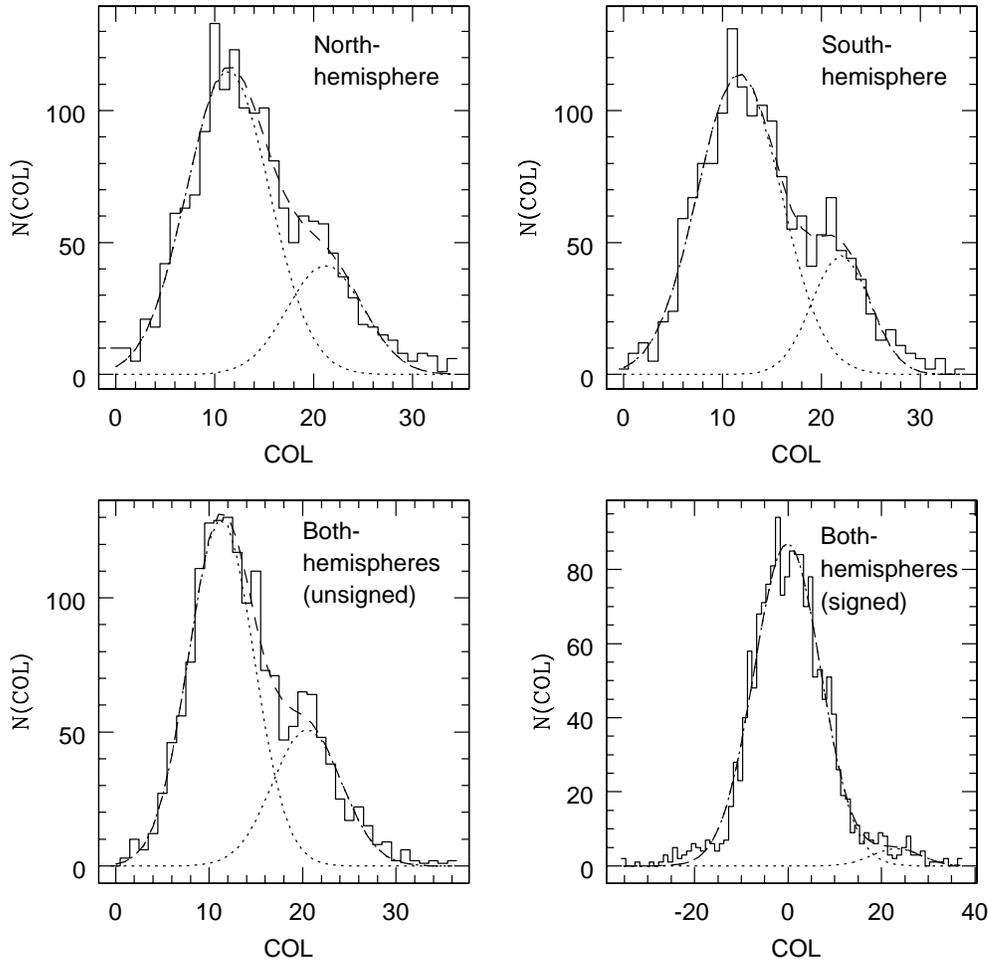}} \caption{Histograms of the COL and their best fits using the double Gaussian function. Solid, dashed, and dotted curves represent the observed histogram, the best fit of the double Gaussian function, and the individual Gaussian function, respectively. In the upper left and right panels, histograms result from sunspots in the northern and southern hemispheres, respectively. In the lower left panel, histograms are obtained by sunspots in both hemispheres using unsigned latitudes of the sunspots. In the lower right panel, histograms are obtained by sunspots in both hemispheres using signed latitudes of the sunspots.}
\end{figure}

\begin{figure}
\resizebox{\hsize}{!}{\includegraphics{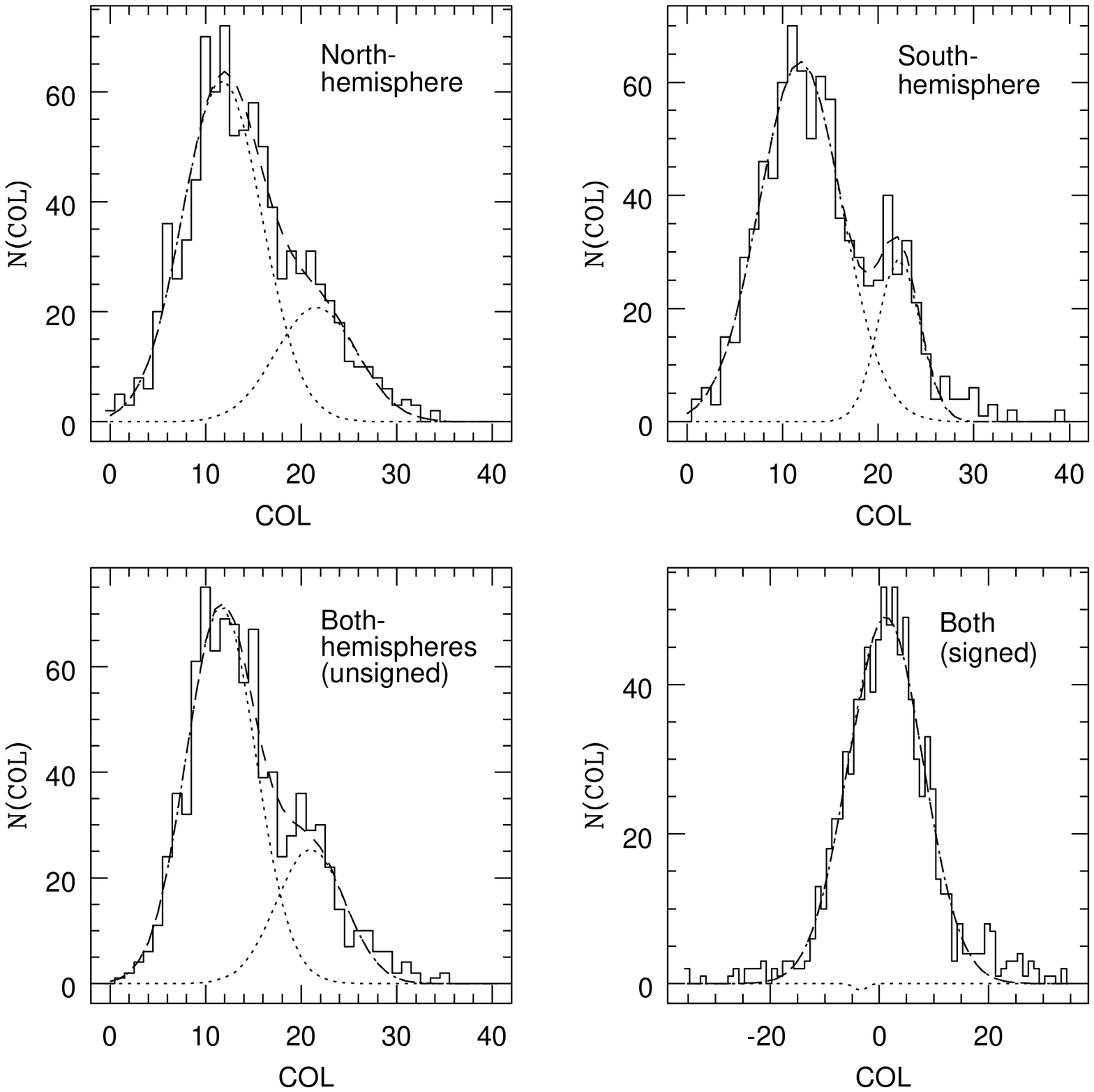}} \caption{Similar to Figure 2, but the histograms are a result of the sunspots observed in the solar cycles when the northern hemisphere is dominant.}
\end{figure}

\begin{figure}
\resizebox{\hsize}{!}{\includegraphics{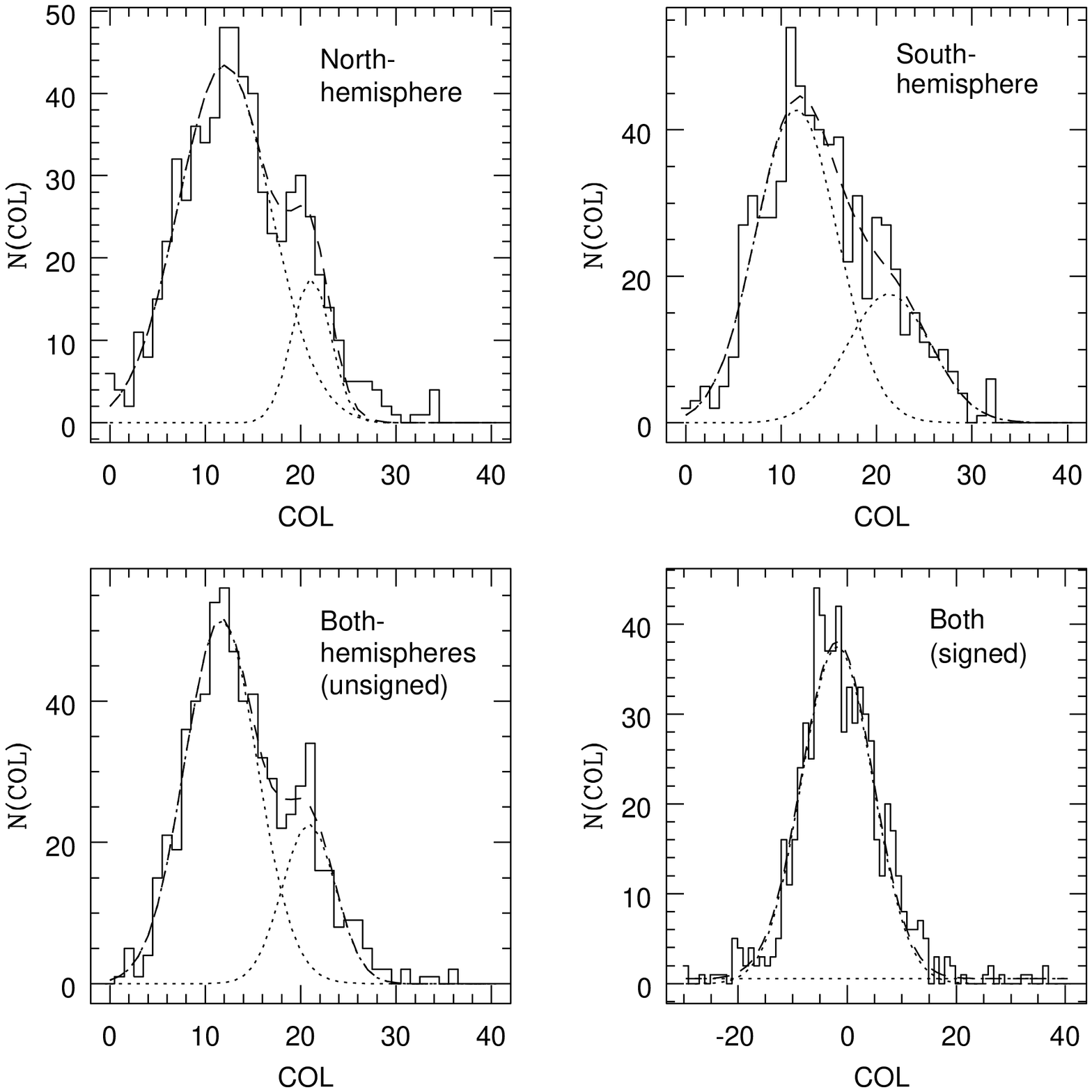}} \caption{Similar to Figure 2, but the histograms are a result of the sunspots observed in the solar cycles when the southern hemisphere is dominant.}
\end{figure}

\end{document}